\documentclass[aps,prl,reprint,groupedaddress,nobibnotes,showpacs]{revtex4-1}

\usepackage{graphicx}
\usepackage{dcolumn}
\usepackage{bm}
\usepackage{natbib}

\begin{document}

\title{A Proposal of a Superconducting Linac in CW Operation for Multi-Purposes}
\author{Miho Shimada}
\affiliation{High Energy Accelerator Research Organization, KEK, Oho 1-1, Tsukuba, Ibaraki, 305-0801, Japan}

\date{\today}

\begin{abstract}
A superconducting (SC) linac is expected to lead to outstanding discoveries in various scientific fields because its beam current is a few orders of magnitude larger than in a normal-conducting linac. However, the widespread use of SC linac is limited by the high construction and operation costs. To resolve this problem, we propose a continuous wave (CW) operation of a SC linac shared by electron/positron beams for effective multi-purposes utilization. A high current positron/electron beam is required for high-energy physics projects such as linear collider and muon collider while high-current and high-quality electron beams is expected to realize the next generation X-ray light sources. As an example, we discuss the injector of the International Linear Collider, an X-ray free-electron laser and an energy-recovery linac light source. We found a feasible solution for the proposed multibeam operation  despite the high-quality beam requirements and complicated opertion: control of mixed beams without pulsed magnets, lower beam loss and heat load in the cavity, high stability of beam energy, and operation at high average current.
\end{abstract}


\maketitle

As the surface of a superconducting (SC) accelerator cavity has an extremely small resistance, accelerating RF fields can be applied with little heating. Consequently, the beam repetition rate can be increased during long-pulse or continuous-wave (CW) operation, and the cost performance per beam current is the highest among linear accelerators such as normal-conductivity cavities, and laser based accelerators. This technique is therefore expected in state-of-the-art large-scale linear accelerators for projects in high-energy particle physics, photon science, and neutron science and applications \cite{ILC-TDR, EuroXFEL, LCLS-II-CDR, ESS-TDR}. Some of these projects utilize the 1.3~GHz TESLA nine-cell cavity developed under the International Linear Collider (ILC) project \cite{ILC-TDR}, an electron--positron collider aiming at the collision energy of 500~GeV. Although it is developed for 1~ms pulse operation, there are some requirements of longer pulses or CW operations even if the acceleration gradient is relatively lower.

As one example, future linear collider projects such as the ILC and the Compact Linear Collider (CLIC) \cite{CLIC-CDR} require high-current positron sources with a flux of $10^{14-15} \ e+/s$, which is several orders of magnitude larger than the existing positron source. The positron flux is expected to be even higher for muon colliders \cite{muon, muonCERN}. Among the most serious technical problems is the thermal loading of the positron target. Obviously, the operation with longer pulse reduces the peak current and mitigates the thermal loading. Positrons can be generated by driving electrons or gamma-rays towards a metal target with a large atomic number. In the baseline of the ILC, positron source is based on gamma-rays from a helical undulator driven by 150~GeV electron beam. Because of the high operation cost of the high-energy electron beam, the pulse duration is limited to 1~ms. Therefore, several positron sources driven by a lower energy electron beam are proposed. One of them is based on gamma-rays by the inverse Compton scattering with lower electron beams, which is potentially be realised the SC linac in CW operation by a technique of energy recovery \cite{M.Shimada-IPAC13, M.Kuriki-AIP08}. Another is called a conventional method, directly driving electron beams toward the target \cite{T.Omori-NIM12,M.Kuriki-LINAC16}. The conventional method is considered as the backup scheme because the required energy of the drive electrons is only a few GeV and it enables the long pulse operation at a reasonable operation cost for reducing the thermal loading.


The SC linac is also a promising facility to realize X-ray light sources with high flux and brilliance owing to the high electron beam current. The European X-ray Free-Electron Laser (XFEL) at DESY \cite{EuroXFEL} was successfully generated the first X-ray laser with a SC linac. Subsequently, SLAC started the LCLS-II XFEL project \cite{LCLS-II-CDR}, which will operate the SC linac in CW mode. As an X-ray source with exceptional brightness, XFEL with an oscillator is also proposed \cite{K.J.Kim-PRL08}. Alternatively, one of the candidates of the next-generation X-ray light source is an ERL composed of a SC linac and a recirculation loop \cite{3GeVERL-CDR, CornellERL-PDDR, R.Hajima-RAST10}. Because the recirculating beam returns its beam energy to the linac, the average beam current is much higher than a linac without the recirculation loop. The km-scale recirculating loop of an ERL provides 20--30 beamlines, whereas the XFEL provides only a few beamlines. 

However, the huge construction and operation costs of the SC linac are critical issues. To conserve these costs, several projects accelerate the electron beam multiple times through the SC linac \cite{CEBAF-IPAC16,CBETA-IPAC17,C.E.Reece-PRAB16,PERLE-CDR}. In this Letter, we propose another approach: effective use of a SC linac by sharing of multi-purpose tasks. As one example, we apply an electron/positron multibeam as the injector of the ILC, XFEL and ERL light source (ERL--LS) with little degradation of the source performances \cite{M.Shimada-IPAC16,M.Shimada-PASJ17}. The multibeam injection have been already proven successful at KEK as the injector for the storage rings of the KEKB accelerator and the Photon Factory \cite{Y.Ohnishi-LINAC06}. However, several features that are not required in KEK multibeam operation may be crucial in this proposed scheme. These features include control of mixed beams without pulsed magnets, low beam loss and heat load in the cavity, high stability of beam energy, operation at high average current (up to 10~mA), and bunch compression.

\begin{figure*}[t]
\includegraphics[width=\linewidth]{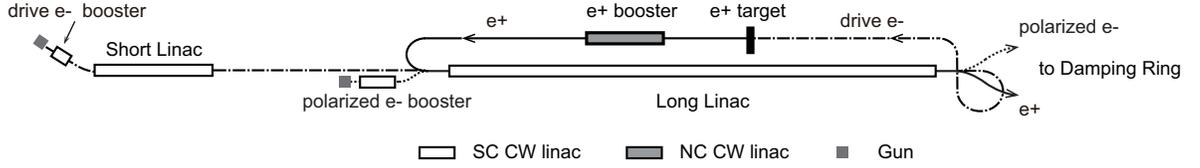}
\caption{\label{ILClayout}Schematic of the proposed layout for the ILC injector with positrons, drive electrons, and polarized electrons.}
\end{figure*}

\begin{figure*}[t]
\includegraphics[width=\linewidth]{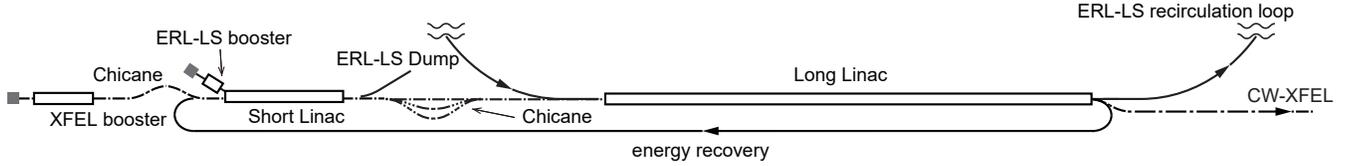}
\caption{\label{LSlayout}Schematic of the proposed layout of the XFEL and ERL light sources. Two chicanes are located between the linacs for individual control of each beam.}
\end{figure*}

As the current backup scheme of the positron source of the ILC is based on a NC linac, the long pulse consists of thousands of micro pulses (duration $1~\mu$s) at 300~Hz to avoid heating by RF fields \cite{T.Omori-NIM12,M.Kuriki-LINAC16}. Owing to the complicated pulse structure, two individual GeV-class NC linacs are required to accelerate the drive electrons for the target and the positrons for injection into the 5~GeV damping ring (DR). The long-pulse structure of the positron is transformed into 1-ms pulses before injection into the main linac.  In addition, the polarized electrons for the ILC collision experiment must be injected into their own DR through a third linac. The three linacs are collectively called the ``ILC injector''. In our proposal, the three beams (polarized electrons, positrons and drive electrons) share a single 6-7 GeV SC injector linac. 
   

The layout of our proposal is shown in Fig.~\ref{ILClayout}. The 6-7 GeV SC injector linac is divided into a short and a long linacs. First, the drive electrons are accelerated sufficiently to hit the positron target. Because the resulting positron is low-quality and accompanied by other radiation particles emitted from the target, it is accelerated by the NC booster to nearly 400~MeV and collimated before injection into the long SC linac to avoid breaking the superconductivity. The beam optics of the long SC linac are optimized such that the transverse size of the positron beam remains much smaller than the iris radius of the accelerating cavity (35~mm), even during multibeam operation. For easy beam operation, the polarized electrons ans positrons are injected at the same energy, while the drive electrons are injected at a higher energy. The energy of each electron beam differs at the end of the SC linac (see Table~\ref{energies}), so the orbits of the three beams can be separated without pulsed magnets. The NC booster should be operated in long- pulse or CW mode, but the frequency can be other than 1.3~GHz. In the VHF region ($\sim$ 180~MHz), long period CW operation has been demonstrated in an NC linac \cite{N.G.Gavrilov-IEEE91}.


\begin{table}[bth]
\centering
\caption{\label{energies} Beam energies at the entrance to the short linac ($E_{in}^{short}$), long linac ($E_{in}^{long}$), and at the exit from the long linac ($E_{out}^{long}$). The electrons for XFEL in the short linac and the drive electrons in both linacs are accelerated at an off-crest phase.}
\begin{ruledtabular}
\begin{tabular}{lccc}
  & $E_{in}^{short}$ [MeV] & $E_{in}^{long}$ [GeV] & $E_{out}^{long}$ [GeV]  \\ \hline
XFEL & 500 & 2.4 & 7 \\
ERL--LS(acc) & 30 & 1.9 & 6.5 \\
ERL--LS(dec) & 1900 & 6.5 & 1.9 \\
Drive e- & $\sim$30 & 1.7 & 5.7 \\
e+ & - & 0.4 & 5 \\
Polarized e- & - & 0.4 & 5 \\
\end{tabular}
\end{ruledtabular}
\end{table}

The SC linac has sufficient capacity to furnish high-brilliance X-ray light sources because the ILC requires only a low average current (less than a few hundred $\mu$A) and a blank interval exceeding 100~ms at 5~Hz for the damping time. As the light source, we consider both XFEL and ERL--LS. The XFEL provides a high-brilliance, large-flux light source whereas the ERL--LS enables a large number of beamlines. The layout of the XFEL and ERL--LS is shown in Fig.~\ref{LSlayout}. The two SC linacs are consistent with Fig.~\ref{ILClayout}. The beam energies at the entrances and exits of the linacs differ by at least 10\% (see Table~\ref{energies}). The energy gain depends on the accelerating phase. The chicane located between the long and short linac enables individual control of the orbit and optics of each beam. Bunch compression of the electrons for XFEL \cite{EuroXFEL} can then be accomplished, by energy chirp induced by off-crest acceleration and non-zero longitudinal dispersion through the chicanes. 

\begin{table}[bth]
\centering
\caption{\label{parameters} Main beam parameters and heat load power in the long linac. Energy recovery (ER) is the summed current of the accelerated and decelerated beams. HR means heat load. }
\begin{ruledtabular}
\begin{tabular}{lccccc}
  & $\varepsilon_n$ [m$\cdot$rad] & q [nC] & $\sigma_z$ [mm] & I [mA] & HL [W/m] \\ \hline
XFEL & $1\times10^{-6}$ & 0.3 & 0.02 & 0.1 & 1 \\
ERL--LS & $1\times10^{-7}$ & 0.01 & 0.3 & 20 (ER) & 5 \\
Drive e- & $1\times10^{-4}$ & 3 & a few & 0.05 & $<1$ \\
e+ & $1\times10^{-2}$ & 3 & a few & 0.05 & $<1$ \\
Polarized e- & $1\times10^{-4}$ & 3 & a few & 0.05 & $<1$ \\
\end{tabular}
\end{ruledtabular}
\end{table}

For the ERL--LS, the design value of the average electron beam current is 10~mA because the threshold current due to beam break up caused by higher order modes (HOM--BBU) is envisaged to be a few tens mA \cite{G.H.Hoffstaetter-PRASTAB04, S.Chen-2016}. In this estimate, the cavity type is assumed as the KEK--ERL model-1 cavity: that is an ILC nine-cell cavity equipped with enlarged beam pipes \cite{H.Sakai-ERL07,N.Nakamura-ERL15}. The designed booster energy for the ERL--LS is approximately 30~MeV to maintain the power consumption at 0.3~MW even without energy recovery. The average current for the XFEL is limited to a few 100~$\mu$A to prevent radiation hazard at the beam dump. The total heat load generated by the beam transport in 2~K region of the long linac \cite{H.Sakai-ERL07,J.Gao-PAC95} is assumed to be approximately 10~W, as shown in Table~\ref{parameters}. The heat load of the short linac is less than that of the long linac.  


\begin{figure}
\includegraphics[width=0.9\linewidth]{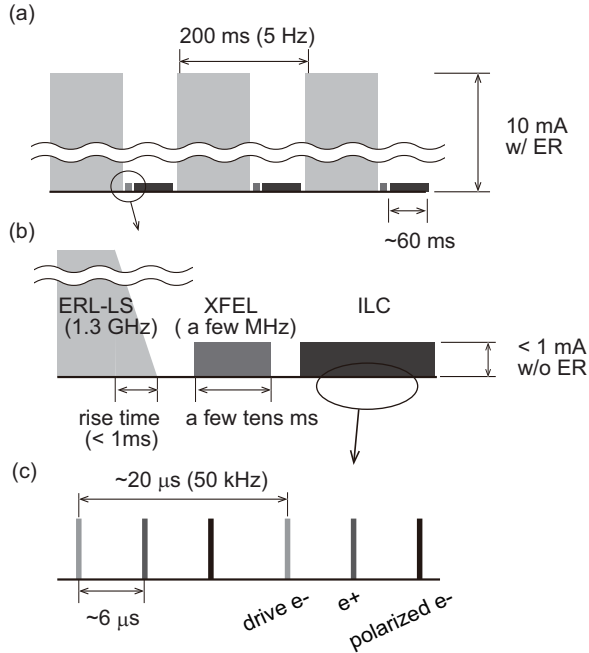}
\caption{\label{PulseStructureALL}Example of the bunch and pulse structure of the electron/positron beams for the ILC, XFEL, and ERL--LS.}
\end{figure}

\begin{figure}[th]
\includegraphics[width=\linewidth]{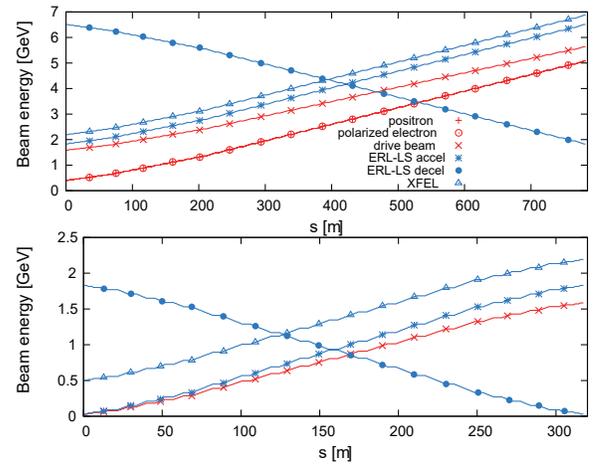} 
\caption{\label{BeamEnergy}Energies of the various beams along the long linac (upper panel)  and short linac (bottom panel). Note that the energies of the positrons and polarized electron beam overlap. }
\end{figure}

As also assumed in the backup scheme, the pulse durations of the proposed ILC injector are assumed to be the 60~ms. Moreover, the bunch repetition rate is maintained constant at 50~kHz, corresponding to 1-ms pulses at 3~MHz in the main linac of the ILC. The long pulses of the positrons and polarized electrons should be overlapped to match the collision timing of the ILC experiments. On the other hand, the positrons trail the drive electrons by almost 6~$\mu$s because they turn around after the almost 700~m-long-linac (see Fig.~\ref{ILClayout}). For simplicity, the polarized electrons are injected behind the positrons with the same delay time. The pulse and bunch structure of the electron/positron beams are shown in panels  (a) and (c) of Fig.~\ref{PulseStructureALL}, respectively. Hereafter these three pulses are collectively termed the ``ILC pulse.''

The bunch charge of 3~nC in the ILC causes beam loading, which (according to rough estimate) induces $O(0.1\%)$ fluctuations of the accelerating field in the cavity \cite{M.Omet-PRSTAB14, T.Schilcher-PHD}. These fluctuations induce correlated electron fluctuations at a much higher repetition rate (a few MHz or 1.3~GHz for X-ray sources). As the electron energy stability of an X-ray light source must be better than 0.01\% the electrons for the light source are operated in the 140-ms blank interval between the ILC pulses to avoid beam loading. In the same way, beam loading of a few hundred pC per bunch in the XFEL induces non-negligible energy fluctuations of the electrons for the ERL--LS. Therefore, the three pulses---ILC, XFEL, and ERL--LS---must never overlap, as shown in Fig.~\ref{PulseStructureALL} (a) and (b). The blank interval time between the three pulses is larger than the time constant of the accelerating cavity (which is of the order of a few ms), sufficient for stabilization of the accelerating field by feedback systems. 

In long-pulse operation, the beams must be accelerated without energy recovery (ER) until the head of the electron beam pulse returns from the recirculation loop. For this reason, the average current of the pulse for the ERL--LS gradually increases during part of the rise time (Fig.~\ref{PulseStructureALL} (b)). The return of the current pulse from the km-scale recirculation loop consumes a few $\mu$s. SC linacs can also accelerate the average electron beam current from 100~$\mu$A to 1~mA without ER. Therefore the rise time was selected from the few tens-to-few hundred $\mu$s range. On the other hand, ER is unnecessaly for the ILC and XFEL because of the low average beam current. 


Each beam can be accelerated or decelerated at phase optimized for each beam. The ILC beams are accelerated in slightly off-crest phase to minimize the effect of the internal wakefield caused by the 3~nC bunch charge, and the electrons for the XFEL are mainly optimized for bunch compression as mentioned above. Meanwhile, the electrons for the ERL--LS should be perfectly on-crest acceleration to minimize the energy spread induced by the 1.3~GHz RF curve.

\begin{figure}[th]
\includegraphics[width=\linewidth]{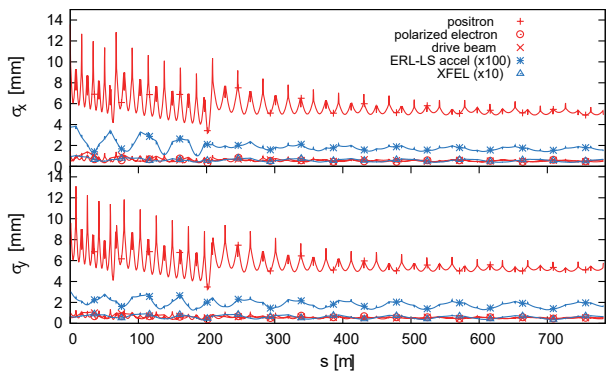} 
\caption{\label{LongLinacSig} Beam sizes in the horizontal (upper panel) and vertical (lower panel) directions of the long linac.}
\end{figure}

The energy of each beam differs along the two linacs, as shown in Fig.~\ref{BeamEnergy}. The accelerating gradient of the SC linac is assumed as 15~MV/m. The full acceleration in the long and short linacs are 4.6~GeV and 1.9~GeV, respectively. The beam-focusing system consists of quadrupole triplets inserted between the SC accelerating cavities. The focusing strength of each quadrupole magnet is inversely proportional to the beam energy. Under this condition, the entire beam optics for transporting the beams at a reasonable transverse size are determined by the following strategies.

The maximum allowable beam loss in the cavity (in other words, in the 2~K region) is assumed as 1~nA/m, corresponding to a few W/m for a multi-GeV beam. Therefore the linear optics are designed so that the root-mean-square (rms) transverse size of the positron beam in the cavity, $\sigma$, is sufficiently smaller than the iris radius of the accelerating cavity $r$, i. e. $5 \sigma < r$. The transverse beam is easily spread out at lower beam energies. Therefore, more quadrupole triplets are this inserted at lower energy region. On the other hand, the envelope function (betatron function) must be refrained from increasing. The betatron function of the decelerated beam of the ERL--LS, which loses energy in the downstream, must be maintained; otherwise, the threshold beam current due to the HOM--BBU can be reduce. To restrain the envelope function, the focus strength is controled in the downstream cavities. Therefore, the rms transverse beam size of the positron beam is focused to not less than approximately 6~mm (see Fig.~\ref{LongLinacSig}) although adiabatic damping will shrink the actual size. Note that the quadrupole triplet allows transverse beam sizes above 6~mm because it operated at room temperature. By virtue of this strategy, the betatron function of the ERL--LS is suppressed to within 120~m.  Meanwhile, the rms transverse sizes of the electron beams for the XFEL and ERL--LS are maintained at less than 100~$\mu$m and 40~$\mu$m, respectively.

\begin{figure}[th]
\includegraphics[width=\linewidth]{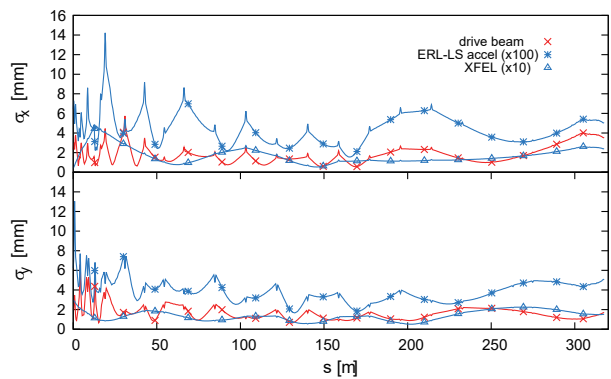} 
\caption{\label{ShortLinacSig} Beam sizes in the horizontal (upper panel) and vertical (lower panel) directions of the short linac.}
\end{figure}

As the linac is divided into two main sections, no positrons pass through the short linac, so the optimal conditions for the ERL--LS (low energy beam at a few tens of MeV) can be achieved. The maximum energy ratio between the accelerated and decelerated beams is high (almost 60), due to the injection and dump energy of 30~MeV and the full energy of 1.9~GeV. Some strategies are necessary to obtain reasonable optics without pulsed magnets. First, the beam optics are symmetrically designed over the entire short linac, because the energies of the accelerated and decelerated beam are symmetric. Second, the periodicity of the quadrupole triplet optics is slightly broken to maintain a small betatron function \cite{I.V.Bazarov-IPAC01}. 
As shown in Fig.~\ref{ShortLinacSig}, the rms transverse beam sizes of the XFEL and ERL--LS are below than 500~$\mu$m and 150~$\mu$m, respectively. If the emittance growth in the recirculation loop is negligible, the rms transverse size of the accelerated and decelerated beam are symmetric (data not shown).

In this Letter, we have proposed a shared SC linac as the ILC injector: polarized electron, positron and their drive electron beams, and high-quality electron beams for the XFEL and ERL--LS as the X-ray lightsource. The electron beams for the light source are operated in the blank interval between the ILC pulses not to be affected by beam loading effects. Although each beam has a different energy, we have now designed reasonable linear optics for the both linacs for the multibeam operation. 

\begin{acknowledgments}
I would like to express gratitude for fruitful discussion of Dr. M. Yamamoto and Prof. K. Yokoya at KEK. I also thank comments of Dr. R. Hajima at National Institutes for Quamtum and Radiological Science and Technology and Prof. M. Kuriki at Hiroshima University.
\end{acknowledgments}


\end{document}